\documentclass{emulateapj}      

\begin{document}

\title{Environmentally-Driven Evolution of Simulated Cluster Galaxies}

\author{Stephanie Tonnesen, Greg L. Bryan, J.H. van Gorkom}
\affil{Department of Astronomy, Columbia University, New York, NY 10027}

\begin{abstract}

Galaxies in clusters are gas-deficient and a number of possible
explanations for this observation have been advanced, including
galaxy-cluster tidal interactions, galaxy harassment, and ISM-ICM gas
stripping.  In this paper, we use a cosmological simulation of cluster
formation and evolution in order to examine this issue from a
theoretical standpoint.  We follow a large number of galaxies over
time and track each galaxy's gas and stellar mass changes to discover
what mechanism(s) dominate the evolution of the cluster galaxies.  We
find that while gas is lost due to a wide variety of mechanisms, the
most common way is via a gas-only stripping event, and the amount of
gas lost correlates with the ram-pressure the galaxy is experiencing.
Although this gas-stripping occurs primarily in the central region ($r
< 1$ Mpc), it is an important mechanism out to the virial radius of
the cluster.  This is due to the wide scatter in ram-pressure strength
that a galaxy experiences at fixed radius.  We find that the timescale
for complete gas removal is $\ge$ 1 Gyr.  In addition, we find that
galaxies in the field and in the cluster periphery ($r > 2.4$ Mpc)
often accrete cool gas; the accretion stops between 1-2.4 Mpc,
possibly indicating the onset of galaxy starvation.

\end{abstract}


\section{Introduction}

It has long been known that a galaxy's environment has an impact on
its morphology: spirals dominate in the field, and ellipticals and S0s
are more prevalent in dense cluster environments (Hubble {\&} Humason
1931).  This has been quantified in the density-morphology relation
(Oemler 1974; Dressler 1980), but this relation alone does not
determine whether the cluster environment affects the formation or the
evolution of galaxies.  According to the Butcher-Oemler effect,
cluster galaxies at z $\ge$ 0.2 are bluer than nearby clusters
(Butcher {\&} Oemler 1978), indicating that cluster age or
evolutionary status may be related to the evolution of galaxies from
spirals to earlier types.  It has been found that the spiral to S0
ratio increases with decreasing redshift (Dressler et al. 1997; Fasano
et al. 2000).  More recently, by examining a specific merging cluster,
Tran et al. (2005) presented a strong relationship between the
Butcher-Oemler effect and infalling galaxies.

More generally it is found  that galaxies
evolve both morphologically and spectroscopically, where the fraction of early
type galaxies and non-star-forming galaxies increase with time at a rate
that depends on the local density (Smith et al 2005; Postman et al 2005;
Poggianti et al 1999; Dressler et al 1997).  The timescale for the 
transformation differs for the two processes. Spectroscopic transformation
precedes morphological transformation (Dressler et al 1997, Poggianti et
al 1999). 

 A wide variety of  physical processes may be responsible for these
evolutionary trends.  
Galaxy-intercluster
medium (ICM) interactions are ones in which the gas in a galaxy
interacts with the ambient intercluster hot gas.  One such process is
ram pressure stripping (RPS), which removes the interstellar medium
(ISM) of a galaxy as it moves through the ICM (Gunn {\&} Gott 1972);
ram pressure could also compress the gas within a galaxy to cause a
burst of star formation that would consume gas that has not been
stripped (Fujita {\&} Nagashima 1999).  An ISM-ICM interaction is the
only type of interaction that does not effect the stellar component of
a galaxy.  Starvation,
the removal of the outer gas envelope by the ICM, can result in normal
star formation slowly exhausting the gas reservoir in the central
region of the galaxy (Larson, Tinsley \& Caldwell 1980).  A galaxy can also interact with the cluster potential,
which can strip both gas and stars, or compress the gas to cause an
increased star formation rate (Byrd {\&} Valtonen 1990).   These
galaxy-cluster interactions affect both the stellar and gas components
of galaxies.  There are also possible galaxy-galaxy interactions that
can take place within a cluster.  These include mergers between
galaxies with low relative velocities, as well as galaxy harassment --
high-speed interactions between cluster galaxies (Hashimoto et
al. 1998; Bekki 1999; Barnes {\&} Hernquist 1991; Bekki 1998; Moore et
al. 1996).  These interactions can cause an increased star formation
rate and will also effect both the stellar and gas components of
galaxies.

It is likely that all of these processes occur in clusters.  Attempts
to differentiate the relative importance of these effects have been
widespread through the years.  For example, Bahcall (1977) showed that
X-ray luminosity is positively correlated with the S0/spiral ratio in
clusters (Bahcall 1977), and that the ratio of spirals to S0s
increases radially (Melnick {\&} Sargent 1977).  Observational studies
of HI deficiency have shown that spirals in clusters have less neutral
atomic hydrogen than galaxies of the same morphological type in the
field (see the reviews by Haynes, Giovanelli, {\&}
Chincarini 1984). On the other hand the CO content does not seem
to depend on environment (Stark et al 1986; Kenney and Young 1989).
H~I imaging of spirals in the center of Virgo shows smaller H~I disks than 
stellar disks, pointing to
an ISM-ICM interaction (Cayatte et al 1990; Warmels 1988). More recently
Koopmann and Kenney 2004 have shown that in Virgo the reduced massive star
formation rate is primarily caused by truncation of the star forming disks,
thus it is the removal of the lower density atomic gas that seems to control
the star formation rate.
  Solanes et al (2001) studied HI deficiency in a
sample of 18 cluster regions, and found that HI deficiency decreases
smoothly out to large projected distances from cluster centers. In a recent
HI imaging study of Virgo, Chung et al (2007) find
a number of long one-sided H~I tails
pointing away from the cluster center. These galaxies are likely falling in
for the first time and gas is indeed already being removed at large projected 
distances from the cluster center.   

Others have combined simple analytic models with cluster galaxy
observations to determine the evolutionary mechanism at work.  For
example, Treu et al (2003) use a large wide-field mosaic of HST images
in combination with a simple cluster model to identify the operating
environmental process.  They find that galaxy star formation rate and
morphological type have mild gradients outside of the central Mpc of
the cluster.  This leads them to conclude that only slow ($>$1 Gyr)
processes that can effect galaxies in the outer regions of clusters
could be responsible, and therefore that galaxy starvation and
harassment are the most likely mechanisms that evolve cluster
galaxies.  In addition, detailed investigations of a few individual
galaxies using multiple wavelengths have begun to unravel their
probable histories (e.g. Crowl et al. 2005; Chung et al. 2005).
Observations of NGC 4522 indicate that the galaxy is undergoing ram
pressure stripping, although it is outside of the high density ICM
(Kenney, van Gorkom, \& Vollmer 2004).  Clearly, these varied results
show the complicated nature of galaxy evolution in clusters.

The limitation of all observations is that they are snapshots of a
long process, and cannot make detailed conclusions about both the
history and fate of individual galaxies.  To overcome this limitation,
simulations are now being used to inform the debate about which
morphology-changing mechanism is operating by modeling previously
observed galaxies (e.g. Vollmer, Huchtmeier, {\&} van Driel 2005).
For example, Vollmer (2003) shows that one Virgo cluster galaxy in
particular has both undergone ram pressure stripping {\it and} been
involved in a gravitational interaction.  Others have studied
idealized cases of galaxies in clusters (e.g., Quilis, Moore \& Bower
2000; Roediger 
{\&} Br{\"u}ggen (2006))

In this paper, we take a different approach and use a cosmological
hydrodynamics simulation in order to track the evolution of a large
number of galaxies in the close vicinity of a large galaxy cluster.
The simulation is performed in a cosmological context and includes
dark matter, gas-dynamics and a treatment of star formation (see \S
\ref{sec:sim}).  In this way, we can study the environmental impact of
the intracluster gas, the cluster potential, as well as other galaxies
in a self-consistent fashion.  Although we have insufficient
resolution to determine the morphological type of our galaxies with
confidence, we can compare the gas content of our simulated galaxies
directly to observations.

The advantage of using a cosmological simulation is that we do not
need to guess the specific environmental conditions of our galaxies,
instead the local environment is naturally modeled by the simulation.
This might be important if, for example, the intracluster medium has
internal motions which increase the efficiency of ram-pressure
stripping above that expected in a static medium, or if galaxies
preferentially enter the cluster in groups.  The output from this
simulation can tell us about the evolution of a particular galaxy and
give an overview of what mechanisms are at work in a cluster
environment.  In this paper we concentrate on the latter by following
the changing gas and stellar mass of 132 galaxies in our cluster.  In
particular, we focus on the period after the cluster has been
established and ask how the gas content of infalling galaxies changes.
Our data tell us how the gas content of gas-rich galaxies is affected
by the cluster, an important part of the evolution of spiral galaxies
to earlier types.

This paper is organized in the following way.  After a brief
introduction to our code, we provide a comparison of the general
characteristics of our simulation to current observations (\S
\ref{sec:sim}).  We then explain the construction of our sample of
galaxies and simulated ``observations" (\S \ref{sec:sample}).  In \S
\ref{sec:results} we describe our results and present an analysis of
the importance of the various gas removal mechanisms in our
simulation.  We conclude (\S \ref{sec:conclusions}) with a discussion
of the limitations and implications of our results.

\section{Methodology}

\subsection{Simulation}
\label{sec:sim}

We have simulated a massive cluster of galaxies with the adaptive mesh refinement (AMR) code {\it Enzo}.  This cosmological hydrodynamics code uses particles to evolve the dark matter and stellar components, while using an adaptive mesh for solving the fluid equations including gravity (Bryan 1999; Norman \& Bryan 1999; O'Shea et al. 2004).  The code begins with a fixed, static grid and automatically adds refined grids as required in order to resolve important features in the flow (as defined by enhanced density).

The cluster forms within a periodic simulation box which is 64 $h^{-1}$ Mpc on a side, in a flat, cosmological-constant dominated universe with the following parameters: $(\Omega_0, \Omega_\Lambda, \Omega_b, h, \sigma_8) =$ $(0.3,0.7, 0.045, 0.7, 0.9)$.  We employ a multi-mass initialization technique in order to provide high-resolution in the region surrounding the cluster, while evolving the rest of the box at low resolution.  Timesteps are also refined in the more dense regions to follow in detail the rapidly changing conditions.  The dark-matter particle mass is $6.4 \times 10^{8}$ M$_\odot$, with a gas mass resolution about five times better than this. The whole cluster has about one million dark matter particles within the virial radius, and a typical $L_*$ galaxy is resolved by several thousand dark matter particles.  The adaptive mesh refinement provides higher resolution in high density regions, giving a best cell size (resolution) of 3 kpc. This is sufficient to resolve the large galaxies in which we are interested, and also, we believe, to approximately reproduce effects such as ram-pressure stripping (although it is clear that the internal dynamics of galaxies will not be well resolved).  A study of the gas-stripping properties of this code is presented in Agertz et al. (2006), which demonstrates that the resolution required to correctly reproduce stripping in grid-based codes is less stringent than in particle-based codes.   We discuss our tests of different resolutions in greater detail in Appendix A.  

\begin{figure}
\centerline{\includegraphics[scale=0.4]{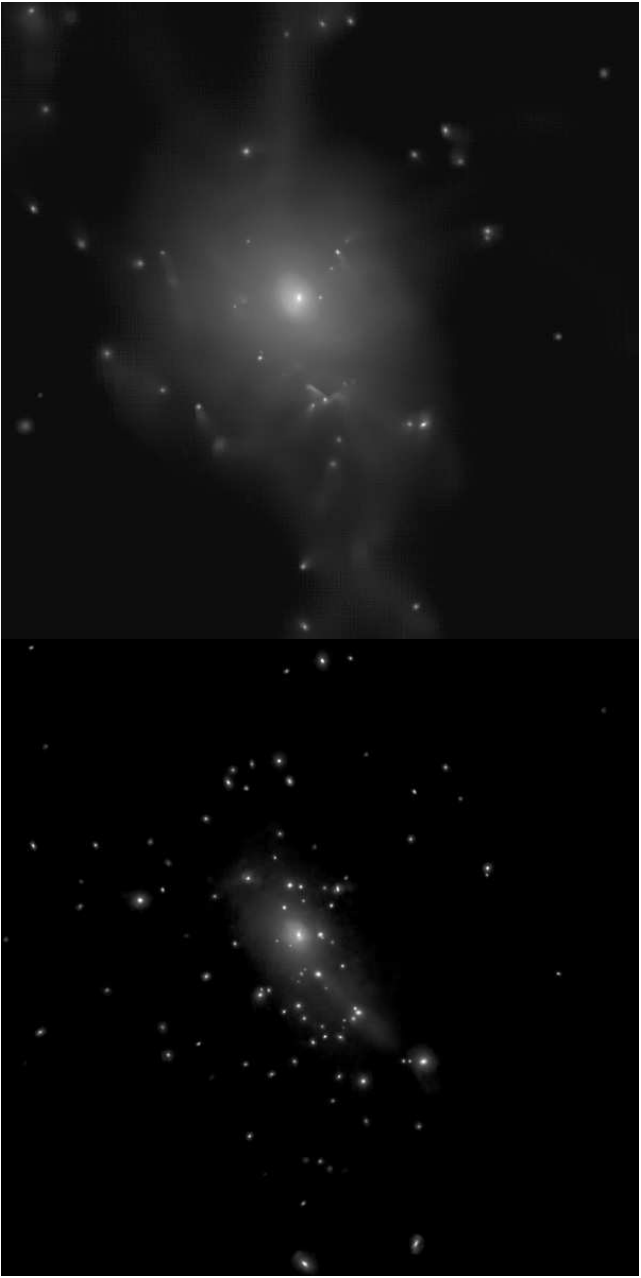}}
\caption{This image shows a projection of the gas density (top) and stellar density (bottom) at z=0.23 in our simulation.  A number of galaxies exhibiting tails of stripped gas can be seen in the gas distribution.  The region shown is 5.1 comoving Mpc on a side, and in both cases the surface density stretch is logarithmic.}
\label{fig:gasimage}
\end{figure}

In Figure~\ref{fig:gasimage} we show a snapshot of the gas and stellar distribution in our cluster.  This shows two features, the first is that most stellar systems within the cluster show only a stellar component, and the second is several clear cases of gas being stripped from galaxies.  These can be seen by the associated tail which often points away from the cluster center.

The simulation includes radiative cooling using the White \& Sarazin (1987) cooling curve, and an approximate form of star formation and supernovae feedback following the Cen \& Ostriker (1992) model.  Briefly, the star formation method relies on identifying cold, collapsing, high-density clouds and forms stars at a rate proportional to the density of gas divided by the dynamical time, multiplied by an efficiency factor.  This efficiency is taken to be, somewhat arbitrarily, 2\%.  See O'Shea et al (2004) for a more complete discussion of the star formation
algorithm.    Stars are represented as stellar particles, and the energy from Type II SN is returned to the gas in
the form of thermal energy.  Although this energetic output can be important, it is known that much of this energy is deposited in high-density gas where the cooling time is short and so is radiated away. This results in an ``overcooling" problem (e.g. Balogh 2001), and manifests itself in our simulations as somewhat overly massive galaxies, as well as a higher-than-observed ratio of stars and cool gas to hot gas.  In addition, we observe hot intracluster gas cooling onto the centers of a few of the most massive galaxies (and in particular the central galaxy).  This cooling is not observed in real clusters, probably because of feedback from supermassive black holes, which are not included in the simulation.

\subsection{Construction of the Sample}
\label{sec:sample}

Although our simulation runs from z = 40, we only wish to compare our results to nearby, virialized galaxy clusters and thus only consider our simulated cluster from z = 0.352 to the present\footnote[1]{Visualizations of these simulations can be found at http://www.astro.columbia.edu/$\sim$gbryan/ClusterMovies}.  We output information from the simulation at time intervals of approximately $0.122$ Gyr (although the timesteps within the simulation are orders of magnitude smaller), for a total of 33 output times.  The output includes (i) the position, mass, size, creation time and metallicity of each star particle, and (ii) the position, mass, cell size, temperature, metallicity, and velocity of each gas cell.

In the construction of our sample, we first separate our star particles into distinct galaxies based on regions of high-density in our N-body stellar code.  A visual inspection of the data shows that (as in real clusters), galaxies are easy to identify because they are highly concentrated, with relatively few stars between galaxies.  This is unlike the case for dark matter substructure, where it can often be quite difficult to associate a given dark matter particle with a given sub-halo.   We used the HOP algorithm (Eisenstein {\&} Hut, 1998), which uses a two-step procedure to identify individual galaxies. First, it assigns a density to each star particle based on the distribution of the surrounding particles and then hops from a particle to its densest nearby neighbor until a maximum is reached. All particles (with densities above a minimum threshold, $\delta_{\rm outer}$) that reach the same maximum are identified as one coherent group.  In the second step, groups are combined if the density at the saddle point which connects them is greater than $\delta_{\rm saddle}$.  We chose HOP because of its physical basis, although we expect similar results would be found using a friends-of-friends halo finder.   We set $\delta_{\rm outer}$,  the minimum density for a particle to be part of a group, to 10000; $\delta_{\rm peak}$,  the minimum central density for a galaxy, to 30000; and $\delta_{\rm saddle}$, the boundary density needed to merge two groups, to 25000 (all density values are relative to the cosmic mean). We chose these values because by visual examination we found that they picked out a single galaxy as the central object; however, reasonable variations in these parameters did not make a significant difference in the number of galaxies.  Using this algorithm, we find that each output (from z = 0.35 until z = 0) has between 155 and 186 galaxies within a 12$^3$ Mpc$^3$ box.

HOP separates the cluster into a set of galaxies at each output, but we still need to identify and follow a set of individual galaxies as they move through the cluster with time.  We expect that the particles with the highest density correspond to the most central particles in a galaxy.  Therefore, we first used the density for each star particle as calculated by HOP to select the 150 densest particles in every galaxy in every output (using the 80 densest particles produced similar results).  For any galaxy with less than 150 particles we used the entire set.  Then, starting with each galaxy in our first output, we looked through the sets of densest star particles in all of the galaxies in the next output for a match between particle identification numbers (which indicates the same star particle at the center of both galaxies).  If there were one or more matches between galaxies in consecutive times, we concluded that we were following a single galaxy.  We continued this process from one output to the next for all 33 times.

Because we intend to compare our results to observations of nearby virialized clusters, we only follow galaxies that were in our box from z = 0.35 until z=0.  There were 155 galaxies in our earliest output, which dictated the maximum number of galaxies we could follow.  Any galaxy that had no match from one output to the next was dropped from our sample.  If a galaxy was dropped at any time, no part of its evolution is reported in the statistical results given below.  Of the 155 galaxies with which we began, we were able to track 133.  Because one of these is the cD galaxy, we report on the evolution of 132 galaxies.  Ten of these galaxies merged before z=0, so by the end of our simulation we report on 126 individual galaxies.  It is notable that although we can track arbitrarily small galaxies, none of the galaxies on which we report ever have less than 150 stellar particles.  In fact, 64 of the galaxies we follow always have at least 3000 particles, and using only those galaxies in the following analysis gives similar results.

Our galaxies were dropped for a number of possible reasons.  They may have been swallowed by the cD, ripped apart by either the cluster potential or galaxy harassment, or merged into a galaxy that we did not follow.  Also, a galaxy was dropped if it left the box.  HOP may erroneously group two distinct galaxies together during a close fly-by, and we dropped a galaxy if it was grouped with one that we did not follow.  By examining our data on the galaxies before they were dropped, we found lower limits for three of our dropping mechanisms:  4 galaxies left the box, 4 galaxies were swallowed by the cD,while 3 galaxies were incorrectly grouped by HOP (out of a total of 22 dropped galaxies).

\subsection{Galaxy Definition}

By following these 132 galaxies, we are able to determine both how many galaxies in a cluster underwent environmentally-driven evolution as well as the mechanisms that were driving their evolution.  To do this, we measured the gas mass and stellar mass of each galaxy through time, using any changes to identify the mechanisms at work.  The mechanisms we consider fall into three broad categories: ISM-ICM, galaxy-galaxy, and galaxy-cluster.  For example, an ISM-ICM interaction would not change the stellar mass of a galaxy, but would reduce the gas mass.  A galaxy-galaxy or galaxy-cluster interaction would effect both the gas mass and stellar mass, and depending on whether the masses increased or decreased, we would label the acting mechanism either tidal accretion or stripping.

Because our determination of the environmental mechanism at work depends solely on the mass evolution of a galaxy, we examined our data in detail to carefully define the boundaries of each galaxy.  Although we were able to use HOP to find our galaxies, locate their centers (defined as their points of maximum density), and track them through time, actually using HOP to determine the stellar mass led to substantial fluctuations in their estimated masses.  To minimize this sort of noise, we defined our galaxies masses to include all the stars and cool gas (T $\le$ 15,000 K) within a sphere of a uniform radius.  We originally chose the largest radius as defined by HOP (90 kpc) and applied it to all galaxies, but found that this introduced unphysical fluctuations in both the stellar and gas mass because it tended to combine distinct galaxies that happened to have overlapping radii.  Therefore we examined the radial density profiles, and found that a radius of $26.7$ kpc almost never included gas from nearby galaxies, while by this radius the gas densities had usually gone to zero.  All but four galaxies were sufficiently compact that the $26.7$ kpc radius included 80\% or more of the mass within the larger 90 kpc sphere.

\section{Results}
\label{sec:results}

\subsection{Gasless Galaxy Distribution}

We first compare the morphology distribution in the simulation and observations.  Because the internal structure of our galaxies is not well-resolved, we adopt a different classification scheme than traditional morphological type.  Instead, we relied on the amount of gas mass, dividing them into two types:  those with and those without cool gas (T $\le 15,000$ K).  We cannot comment on the molecular gas in the simulated galaxies because gas in our simulation cannot radiatively cool to molecular cloud temperatures, and long before it collapsed to the density of molecular clouds it will have formed stars.  Therefore we do not comment on molecular gas throughout this paper, although we do expect that it might be stripped more slowly than the lower density gas that we do follow. 

While this classification scheme is overly simplified, it is still instructive to consider the resulting distribution.  For example, the 132 galaxies that we track through time show a clear cluster-radius and gas content relation, seen in Table~\ref{tbl-1}.  Gasless galaxies begin to dominate the population at 2 Mpc and galaxies with gas dominate at large radii and outside the cluster.  We considered our output for the 132 galaxies at 15 equally spaced ($\approx 0.244$ Gyr) output times, thus giving us 1,980 ``observations".  We treat each data point as distinct because, although every galaxy is measured 15 times, at each observation it is in a different part of the cluster and therefore could be influenced by different environmental effects.  This is less true for galaxies far from the cluster center.

\begin{table*}
\begin{center}
\caption{Simulated Gasless Galaxy fraction compared to T2003 E+S0 fraction\label{tbl-1}}
\begin{tabular}{c | rrrr}
\tableline
 & Inner 200 kpc & Central Region & Transition Region & Periphery\\
 & (0-200 kpc) & (0.2-1 Mpc) & (1-2.4 Mpc) & (2.4-5 Mpc) \\
\tableline
Gasless Galaxy Fraction\tablenotemark{a} & $93.5\% $ & $66.7\%$ & $48.9\%$ & $14.5\%$ \\
E+S0 Fraction\tablenotemark{b} & $75\%\pm10\%$ & $50\%\pm7\%$ & $51\%\pm7\%$ & $42\%\pm13\%$ \\
\tableline
\end{tabular}
\tablenotetext{a}{Fraction of gasless galaxies whose average distance in a timestep is within the stated boundaries.}
\tablenotetext{b}{From T2003 and references therein}
\end{center}
\label{table:fractions}
\end{table*}

Ideally, we would compare our galaxies to observations of the galaxy gas population, however, large, complete samples extending to large radius do not yet exist.  Although our gasless galaxies could be either E+S0s or spirals, we expect the majority of our gasless population to be E+S0s.  If we make this rough translation, we can compare our galaxy distribution to the observations of Treu et al. (2003, henceforth T2003).  We choose this comparison because our simulated cluster has a virial radius (calculated as $r_{200}$) similar to that of Cl 0024+16 (1.8 Mpc and 1.7 Mpc, respectively), and we can split our cluster into the same diagnostic regions as in T2003: 0-1 Mpc is the central region; 1-2.4 Mpc is the transition region; and 2.4-5 Mpc, the periphery.  However, our simulation differs from observations in that we have the exact distance from cluster center of all our galaxies in three dimensions, so we do not have any uncertainty due to projection effects.

In our simulation, gasless galaxies dominate the population within 200 kpc, comprising 93.5\% of the galaxies.  We compare our gasless fraction to the E+S0 fraction reported by T2003 in Table~\ref{tbl-1}.  Within 1 Mpc, our gasless galaxy fraction is larger than the E+S0 fraction of T2003.  Between 1 Mpc and 2.4 Mpc the fraction of gasless galaxies in our simulation is within one sigma of the observed E+S0 fraction.  In the outer region (2.4-5 Mpc) the gasless galaxy fraction is less than half of the fraction of E+S0 galaxies observed in the field.

While this qualitative agreement is promising, we can speculate about the cause of the observed disagreement.  If our simulated cluster were older than that observed by T2003, we would expect our elliptical fraction in the central regions to be larger than the value they find (T2003 and references therein).  Also, our gasless fraction may be high because we have no dilution of our central region sample due to projection effects.  

However, one of the most important reasons for the mismatch between our gasless galaxy distribution and T2003's E+S0's is that we are comparing two different populations and assuming there is significant overlap.  Because we are counting gasless galaxies instead of E+S0 galaxies, any gasless spirals in the inner regions of our cluster will cause our fraction to be higher than T2003's.  For example, Solanes et al. (2001) have observed highly deficient spirals in the central regions of clusters.  Also, E+S0s are not necessarily gasless in the field (Morganti et al. 2006 and references therein).  

In the outer regions, the overabundance of cool gas is consistent with previous simulations of galaxy formation, a problem often referred to as the overcooling problem.  One suggested solution to this problem is energetic feedback from AGN, which can heat up cool gas and eject it from the galaxy (e.g. Bower et al. 2006).  Therefore, we caution that we may be overestimating the number of galaxies with gas and the amount of cool gas some galaxies contain. 

In addition to the overcooling problem in elliptical galaxies in general, one or two of the galaxies in the central regions of our cluster have a very large amount of cool gas ($> 10^{11} M_\odot$).   As noted above, the lack of AGN feedback in our simulations is one possible reason for these high gas masses.  Indeed,  half of the observations of galaxies with gas mass over $10^{11}$ $M_\odot$ are of the largest galaxy we follow at different times.  This galaxy may be large enough that it is susceptible to the same overcooling problem as the cD galaxy. 

\subsection{Simulated Galaxy Evolution as a Function of Environment}

\begin{figure*}
\plotone{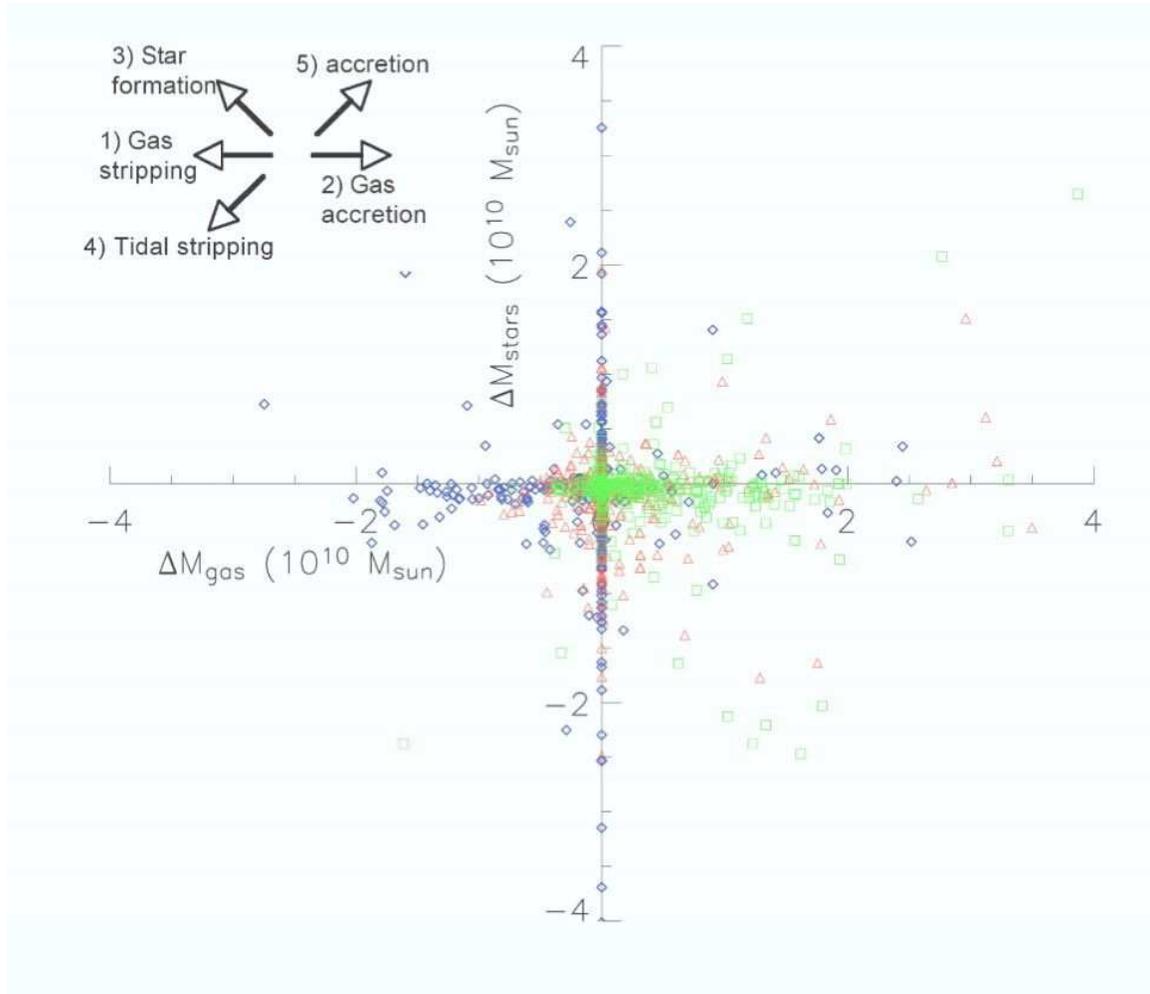}
\caption{Comparison of the change in cool gas mass (T $\le$ 15 000 K) and the change in stellar mass of all observations of galaxies within 5 Mpc of the central cD (1,275 points).  The colors and shapes denote distance to the cD:  blue diamonds are at 0-1 Mpc, red triangles at 1-2.4 Mpc, and green squares are at 2.4-5 Mpc. We have removed the effect of star formation within each galaxy in order to focus on direct environmental effects.}
\label{fig:mass_change}
\end{figure*}

\indent We will now use our simulated sample to examine how a galaxy gains and loses mass.  Recall that we track the evolution of galaxies purely by following the changing gas and stellar mass of each galaxy.  Using only the change in stellar mass and gas mass, we have five likely evolutionary tracks: (i) gas mass loss without stellar mass loss, indicative of an ISM-ICM interaction; (ii) gas mass gain without stellar mass gain, suggesting gas accretion; (iii) gas mass loss equal to stellar mass increase, implying star formation; (iv) both gas and stellar mass loss, indicative of tidal stripping or galaxy harassment; and (v) increase in both gas and stellar mass, suggesting accretion or a merger.  Each of these processes is represented by a unique vector in the ($\Delta M_{gas}$, $\Delta M_{star}$) plane, which is shown in Figure~\ref{fig:mass_change}.

From our simulated data we can compute the change in gas and stellar mass for each galaxy during each timestep ($\Delta M_{gas}$, $\Delta M_{star}$) and plot this value for each galaxy for each time step in Fig.~\ref{fig:mass_change}.  Out of a total of 1,980 points, we only show the galaxies that may be affected by the cluster environment, defined as being closer than 5 Mpc to the cD during each 0.244 Gyr timestep.   This leaves us with 1,257 observations of cluster galaxies.  To make most of the points in our plot more readable we have zoomed in on the inner region of our graph.  Thus two of the merger observations are outside of the range of this plot, and a few observations along the y-axis.  Fewer than 2\% of our observations are outside of the plot range in Figure.~\ref{fig:mass_change}.

Star formation may or may not be environmentally induced, so in this initial examination of galaxy evolution, we remove its effect on the gas and stellar mass.  To do this, we first used the creation time of every star particle in our code to identify the amount of stellar mass formed during a timestep.  We then transferred all of a galaxy's stellar mass that was created during that timestep to the gas mass of that galaxy.  Thus, when we plot a galaxy that had only undergone star formation during a timestep in this figure, we will see no change in either the stellar or gas mass.  This process is done separately for each galaxy and each timestep. 
 
Our results are shown in Figure~\ref{fig:mass_change}, with a more quantitative view of the results in Table~\ref{table:changes}.  In the table, for all nine possible galaxy mass permutations ($\Delta M_{star}$ positive, zero, or negative and $\Delta M_{gas}$ positive, zero, or negative), we list: the number of observations in each category, the total change in gas mass summed over all galaxy observations, and the total stellar mass change.  Notice that the placement of the nine boxes visually corresponds to the nine zones in the plot. 

Since we have removed star formation, most of our galaxies are clustered near the origin, having no large change in mass (see also Table~\ref{table:changes}). This is the largest set of galaxies, and contains both galaxies that have no mass variation in a time-step, and those that only form stars out of their own gas (i.e. are evolving without obvious external influence).  For the purpose of Table~\ref{table:changes}, we treat small mass changes (within $\pm 3.16 \times 10^9 M_\odot$ of zero) as constant for both the gas and the stellar masses.  We chose this value by fitting a guassian to the points near the x-axis of our plot, and adopting the 3${\sigma}$ value.  This choice gives 729 observations in the ``constant" category, which is our largest group.

There are also a number of galaxies -- along the negative x-axis -- which are undergoing gas mass loss only:  it seems likely that these galaxies are affected by ram pressure and associated ISM-ICM stripping processes because there is gas loss and no (or very little) change in stellar mass.  As can be seen in Table~\ref{table:changes}, this category represents the most frequent source of gas mass loss and also results in the highest amount of gas mass loss.  We will examine these systems in more detail in section~\ref{sec:rps}.

The galaxies along the positive x-axis are accreting cool gas from their surroundings, or from a larger halo.  This is also a large group of objects, and such accretion is the primary mode for the growth of gas-rich galaxies in our simulation.  We will discuss them in more detail below.

\begin{figure*}
\plottwo{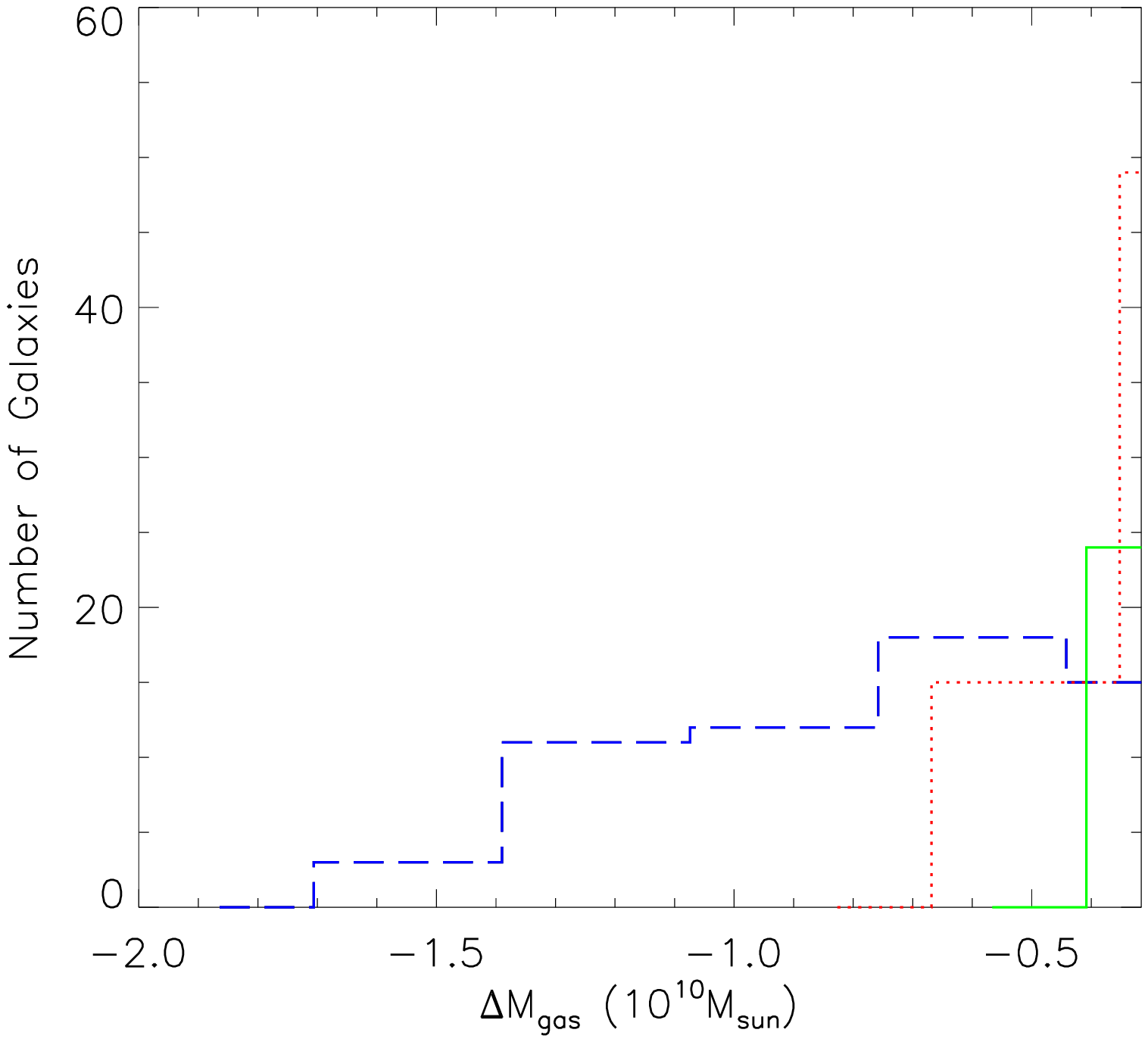}{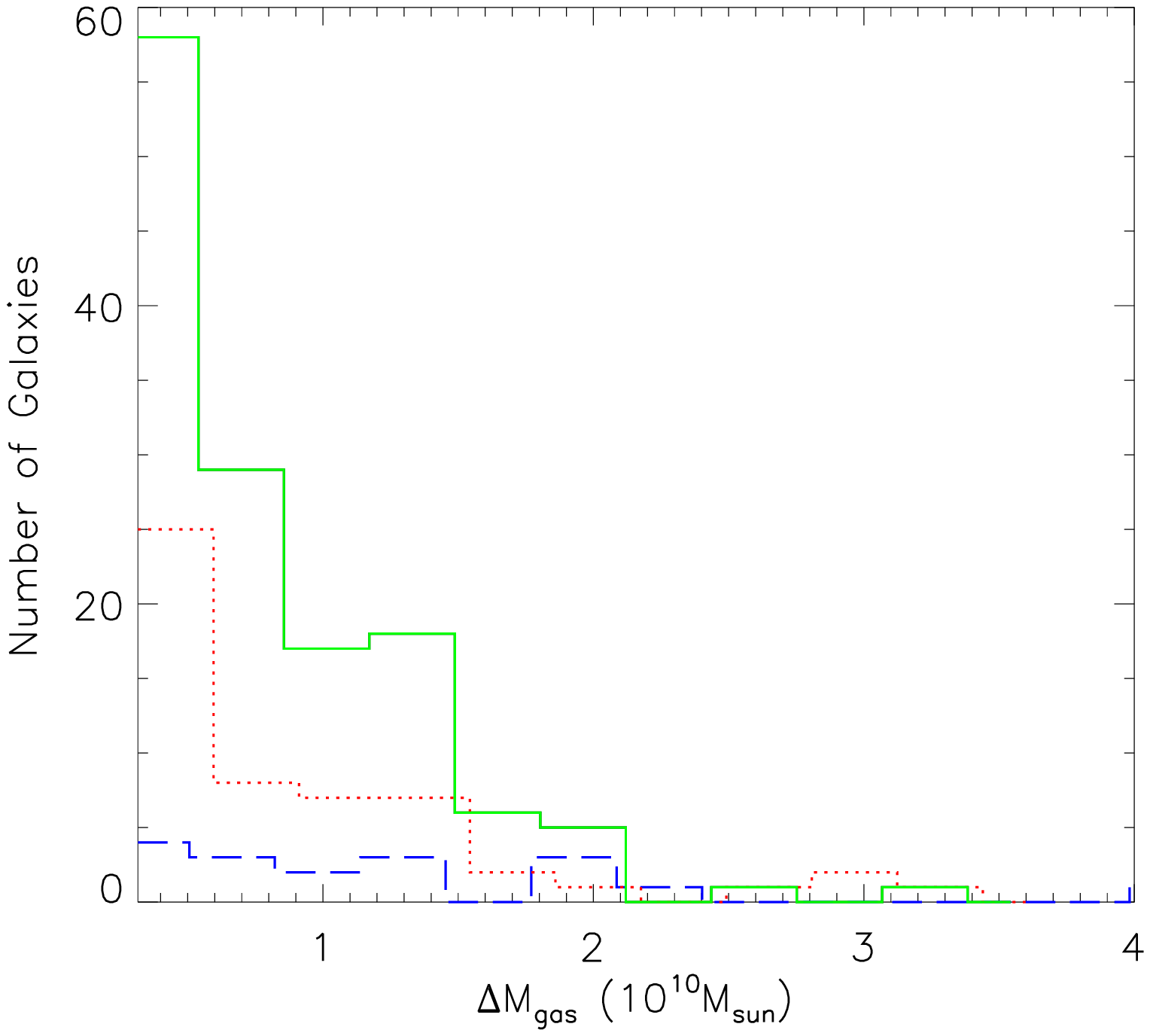}
\caption{ These histograms compare of the changes in cool gas mass (T $\le$ 15 000 K) for the galaxies that have no change in their stellar mass.  The figure on the left records gas mass loss, while the right figure shows the gas mass gain.  The colors denote distance from the cD, as in Figure~\ref{fig:mass_change}.  The central regions galaxies are denoted with a dashed line, the transition region with a dotted line, and the periphery with a solid line.  Note that the central region galaxies are often gasless or undergoing gas stripping.  The transition region galaxies are undergoing more gas stripping than predicted by T2003.  The periphery galaxies are accreting gas as is expected outside of clusters.  Because the large number of galaxies undergoing no environmentally-driven change in gas mass would dominate this histogram, we have cut out a small section surrounding zero gas mass change.}
\label{fig:mass_change_hist}
\end{figure*}

The galaxies that fall along the y-axis are galaxies (most often gasless galaxies with therefore exactly zero $\Delta M_{gas}$) that are either accreting or losing stars.  Some of the points along the vertical axis do not represent true stellar mass gain or loss, but are instead due to double-counting during a near collision.  During close encounters, galaxies first appear to gain stars in one timestep and then lose them in the next as the other galaxy moves into and then out of the spherical galactic region.  This would produce a pair of points with similar magnitudes but opposite signs.  Neither of these points would represent true loss or gain (although harassment might cause some true stellar mass loss).  However, we also expect to see gasless galaxies being stripped of stars by the large cluster potential and possibly by galaxy-galaxy interactions.  This is observed, both in the number of galaxies affected and the total stellar mass gained and lost.  If every instance of increased stellar mass is a product of a close fly-by and thus paired with a stellar mass decrease, we still observe at least 70 galaxies undergoing tidal stripping or harassment.

Besides the axis, the other quadrants are relatively empty.  The galaxies that are observed to lose stars and gain gas (in the lower right quadrant of Fig.~\ref{fig:mass_change}) are likely to be undergoing two processes within one timestep.  When we made the same graph using timesteps half as long we found only $62.5\%$ of our observations remained in the fourth quadrant.  The very small number of galaxies observed to gain stars and lose gas may also be caused by the aggregation of a few processes.  Despite this, we chose not to use the smaller timestep because in most cases it splits up a single process, resulting in more smaller mass changes along the axes.  The points from the central cluster region in the lower right quadrant of Figure~\ref{fig:mass_change} are dominated by the largest galaxy that we follow (7 of 10 points).  These points may be caused by the galaxy being physically stripped by the cD while it is unphysically overcooling surrounding gas.

The galaxies losing both gas and stars may be undergoing galaxy harassment or tidal stripping by either the cD or a nearby galaxy.  In a parallel process, the galaxies that are gaining both gas and stars are either merging or accreting both gas and stars from their surroundings or a nearby galaxy.  We found a lower limit of three mergers by counting the number of tracked pairs that merge within a radius of 5 Mpc from the cD.  It is interesting that all three of these mergers involve galaxies with gas, and all three merged galaxies continue to contain cool gas throughout the simulation.  

The points in Figure~\ref{fig:mass_change} are color-coded by the galaxy's minimum distance from the central cD during each timestep; blue diamonds are within 1 Mpc, red triangles are within 2.4 Mpc, and green squares are out to 5 Mpc.  It is clear from a visual inspection of the plot that many of the central galaxies are either gasless galaxies undergoing a tidal process or galaxies undergoing gas stripping.  The galaxies in the periphery are the majority of the galaxies gaining gas, as spirals are conjectured to do in the field (Larson, Tinsley \& Caldwell 1980).  The transition region galaxies are not so easily categorized, and seem to consist of galaxies undergoing processes more clearly associated with one of the other two regions.  On a qualitative scale, our graph compares well with T2003's Figure 10 in that we see more ISM-ICM interactions close to the cD and mergers spanning the entire 5 Mpc: three of the points in the first quadrant of Figure~\ref{fig:mass_change} are mergers between tracked galaxies, two of which are in the central region and one of which is in the periphery.

In Figure~\ref{fig:mass_change_hist} we look more closely at the points that lie along the positive or negative x-axis, and generate the distribution of gas mass loss and gain, again color-coded by the galaxy distance from the cD.  The left histogram shows that most of the ram pressure stripping occurs in galaxies in the central region, but does also have some impact on galaxies in the transition region and, to a lesser extent, the periphery.  Further, the right histogram shows that gas accretion is occuring, and occurs mostly to galaxies in the periphery.  There is a significant drop in the number of galaxies accreting gas in the transition region in comparison to the periphery, and we may be seeing the region where starvation begins to occur (Larson, Tinsley \& Caldwell 1980).

\begin{table}
\begin{center}
\begin{tabular}{r || c  c  c}
 & Gas Mass & Constant & Gas Mass\\
 &  Loss & Gas Mass &  Accretion\\
 \tableline
 \tableline
 Stellar Mass & 5 & 57 & 26\\
 Accretion & -6.747 & -0.57 & 56.991\\
  & 4.267 & 49.254 & 62.578\\
  \tableline
  Constant & 84 & 729 & 169\\
 Stellar Mass & -65.834 & 6.977 & 158.785\\
  & -5.893 & -27.433 & -4.349\\
  \tableline
 Stellar & 18 & 125 & 44\\
 Mass Loss & -14.83 & -0.312 & 54.597\\
   & -120.412 & -220.414 & -85.618 \\
\end {tabular}
\end{center}
\caption{This charts the possible changes to gas and stellar mass of the galaxies that are observed within 5 Mpc of the cD.  It is a more quantitative description of the information graphed in Figures~\ref{fig:mass_change} and~\ref{fig:mass_change_hist}.  Note that the organization of this chart matches the layout of Fig.~\ref{fig:mass_change}.  For each category we include three rows of information: 1) the total number of observations in that category, 2) the total amount of gas mass lost or gained in all the observations, and 3) the total stellar mass change.  The total mass changes are in units of $10^{10} M_\odot$.  There are a total of 1257 observations.}
\label{table:changes}
\end{table}

\subsection{Ram Pressure Stripping}
\label{sec:rps}

\begin{figure}
\includegraphics[scale=0.5]{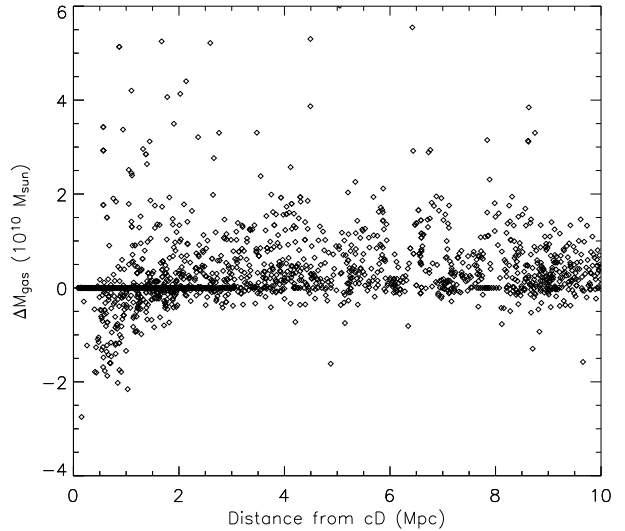}
\caption{The change in gas mass plotted against the average distance from the cD for each timestep.  The amount of gas mass loss per observation and the number of observations of gas loss begin to increase sharply at 2 Mpc, just beyond the $r_{vir}$ (1.8 Mpc) of the cluster.}
\label{fig:gas_loss_dist}
\end{figure}

Of the 107 observations in which a galaxy loses gas by a mechanism other than star formation, at least 84 fit our criteria for ICM-ISM interactions (refer to Table~\ref{table:changes}).  In order to better examine how $\Delta M_{gas}$ changes with radius, we plot $\Delta M_{gas}$ against the distance from the central cD, still removing star formation.   Figure~\ref{fig:gas_loss_dist} shows that there is an increase in both the amount of gas mass lost and the number of galaxies losing gas, with decreasing distance from the cD.  This trend is strongest for $r < 1$ Mpc, but begins at about 2 Mpc, significantly beyond the 1 Mpc radius T2003 had used as the edge of high ICM density.  

\begin{figure}
\includegraphics[scale=0.5]{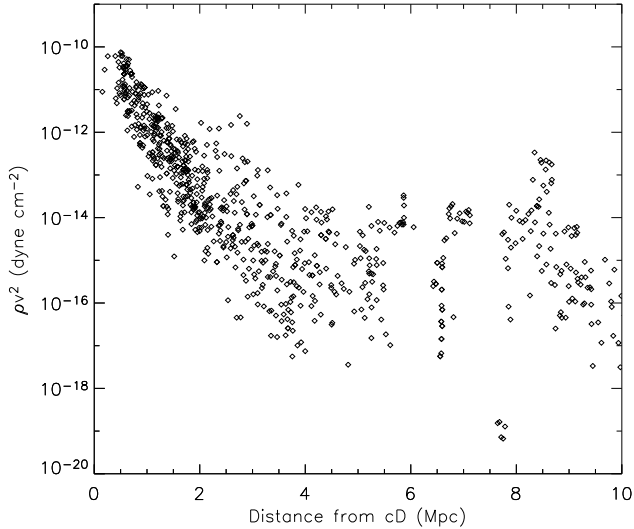}
\caption{Plot of $\rho v^2$ against distance from the cluster center.  In the inner 2 Mpc of the cluster there is an increase in ram pressure, evidence that ram pressure is indeed the cause of the increase in gas loss seen in the galaxies in the inner region of the cluster.}
\label{fig:ram_distance}
\end{figure}

To clarify the reason for this trend, we calculated the ram pressure as first derived by Gunn and Gott (1972):  $\rho v^2$, where $\rho$ is the ICM density and the $v$ is the relative velocity between the ICM and the ISM.  Gas was defined to be part of the ICM if it had a temperature above $5 \times 10^6$ K.  The density was calculated for all the hot gas in a sphere of radius 90 kpc centered on a galaxy center previously identified by HOP.  To find the velocities of the ICM and ISM we averaged the velocities of all the individual cells of gas that were included in the 90 kpc or 26.7 kpc sphere, respectively.  We then took the magnitude of the velocity difference to use in our ram pressure calculation.  Figure~\ref{fig:ram_distance} shows how ram pressure varies with distance from the cD:  there is a definite increase of ram pressure with decreasing distance to the cluster center beginning at about 2 Mpc.  This correlates well with the increasing gas loss with decreasing distance to the cD.  It is also important to note that $\rho v^2$ varies by about two orders of magnitude at a given radius.  This is partially due to the density and velocity structure in the ICM which is apparent in the simulations, and partially due to the wide range in galaxy velocities at a given radius.  To illustrate the importance of this effect, we can see that, for example, at 2 Mpc from the cD, ram pressure is often below the value of $10^{-12}$ derived by Gunn \& Gott (1972) to be the minimum ram pressure for effective stripping, but there are some observations of higher ram pressure values.  

\begin{figure}
\centering
\includegraphics[scale=0.5]{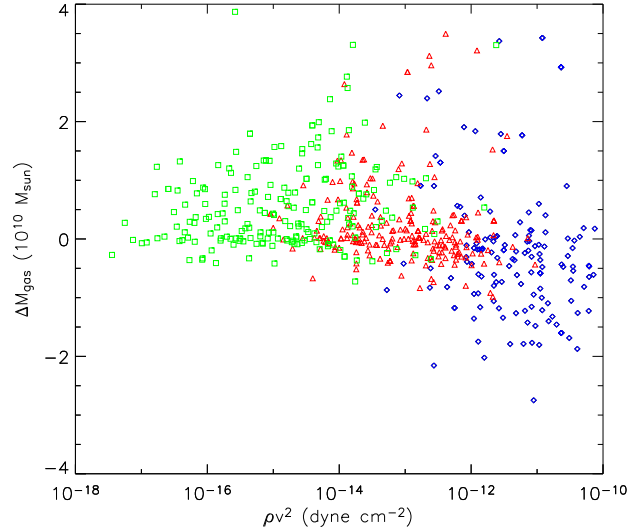}
\caption{A plot of the change in cool gas against ram pressure.  The color-coding is the same as in Figures~\ref{fig:mass_change} and ~\ref{fig:mass_change_hist}.  This also supports the claim that ram pressure stripping is the most important cause of gas loss in these cluster galaxies.  We also see that ram pressure seems to be affecting some galaxies in the transition region (red triangles).}
\label{fig:ram_pressure}
\end{figure}

Figure~\ref{fig:ram_pressure} plots the amount of gas mass loss against ram pressure.  The color-codes are the same as in Figures~\ref{fig:mass_change} and \ref{fig:mass_change_hist}, and in this figure see a definite relation between gas loss and ram pressure.  Although most of the points with a large amount of gas loss and high ram pressure are from galaxies within the inner 1 Mpc of the cluster, there are a few galaxies that seem to be ram pressure stripped from the transition region, consistent with our interpretation of Figures~\ref{fig:gas_loss_dist} and \ref{fig:ram_distance}.

While the correlation with ram pressure strength is indicative of a role for ram pressure stripping, we should note that we do not exclude viscous stripping and other mechanisms which scale in a similar way.  In the following, we refer to these processes collectively as ram-pressure stripping.

\subsection{Case Studies of Gas-stripped Galaxies}

To examine the stripping process in more detail, we examined the 16 galaxies that went from having a cool gas mass of more than $3.16 \times 10^{9}M_{\odot}$ at z $\sim$0.35 to having no cool gas by the end of the simulation.  We chose this mass loss cutoff because it is the limit of our ``constant" category, as discussed earlier.  Of the 119 observations of these 16 galaxies, 37\% fit our gas loss criteria, for a total cool gas mass loss of $3.86 \times 10^{11} M_{\odot}$, caused by all mechanisms (other than star formation).  Of the observations of galaxies losing gas, 89\% have no change in stellar mass, which we take to indicate ram pressure stripping (or a related mechanism).  These observations are distributed among thirteen of the sixteen galaxies, for a total of $3.25 \times 10^{11} M_{\odot}$ gas mass lost by ram pressure stripping. 

To verify that we are not merely seeing a part of a longer episode of tidal stripping or galaxy harrassment that included only gas loss for a subset of the observations, we looked at the four observations in which both gas and stars were lost.  Only one of these was of a galaxy that also contained an observation of pure gas stripping.  Even ignoring this galaxy, 88\% of gas loss observations are of ram pressure stripping.  To be conservative in our number of ram pressure stripped galaxies, we assumed that the galaxy that had only gas loss followed by both gas and stellar mass loss did not undergo ram pressure stripping.  Thus, we only include 12 galaxies in our ram pressure stripping statistics.

Next, we considered where in the cluster these galaxies were being stripped.  Although most of the galaxies undergoing ram pressure stripping were in the central region of the cluster, 40\% of the observations were of galaxies in the transition region.  Of those in the transition region, only 13\% were of galaxies that had been within 1 Mpc of the center since z$\sim$0.35.  Thus, most of our observations of galaxies undergoing ram pressure stripping in the transition region were beginning the ISM-ICM interaction there.  We even observed a single galaxy being ram pressure stripped in the periphery for $\sim2.5$ Gyr.  This galaxy was also unique in that it had lost all of its gas before reaching the central region of the cluster.

Finally, we can give a rough estimate (because our time resolution is 0.244 Gyr) of the length of time it took galaxies to lose their gas once gas loss began.  This is a worthwhile estimate for comparison to T2003, who define any ISM-ICM interaction that is longer than 1 Gyr as starvation.  We find that 5 of the 12 galaxies fulfilling our ram pressure stripping requirement lose their gas in about 1 Gyr.  However, we also find that 5 galaxies lose their gas in well over 1 Gyr, and only 2 galaxies lose their gas in much less than 1 Gyr.  This is in tentative agreement with T2003's conclusion that galaxy transformation is generally a slow process.  
 
\begin{figure*}
\centerline{\plottwo{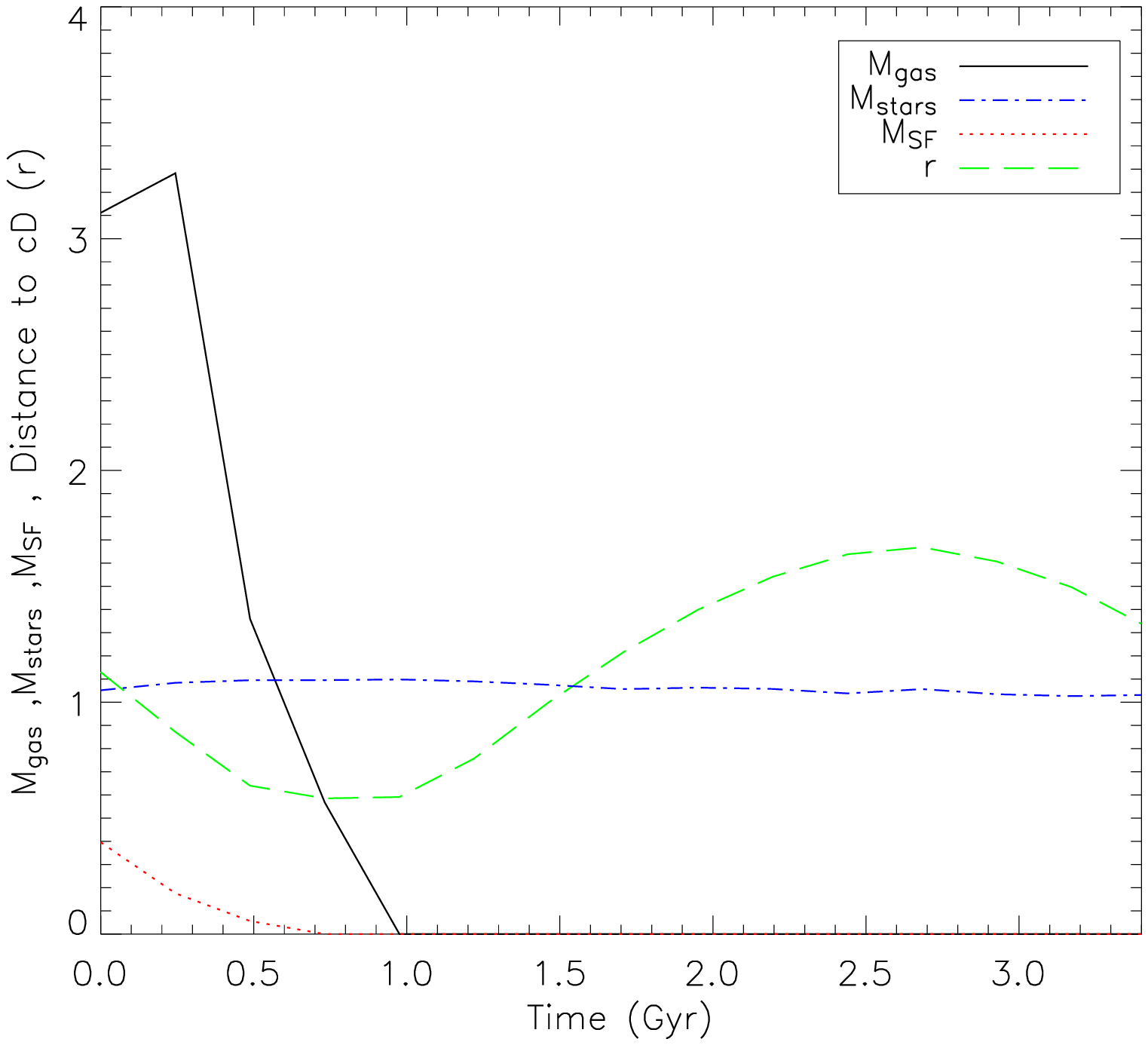}{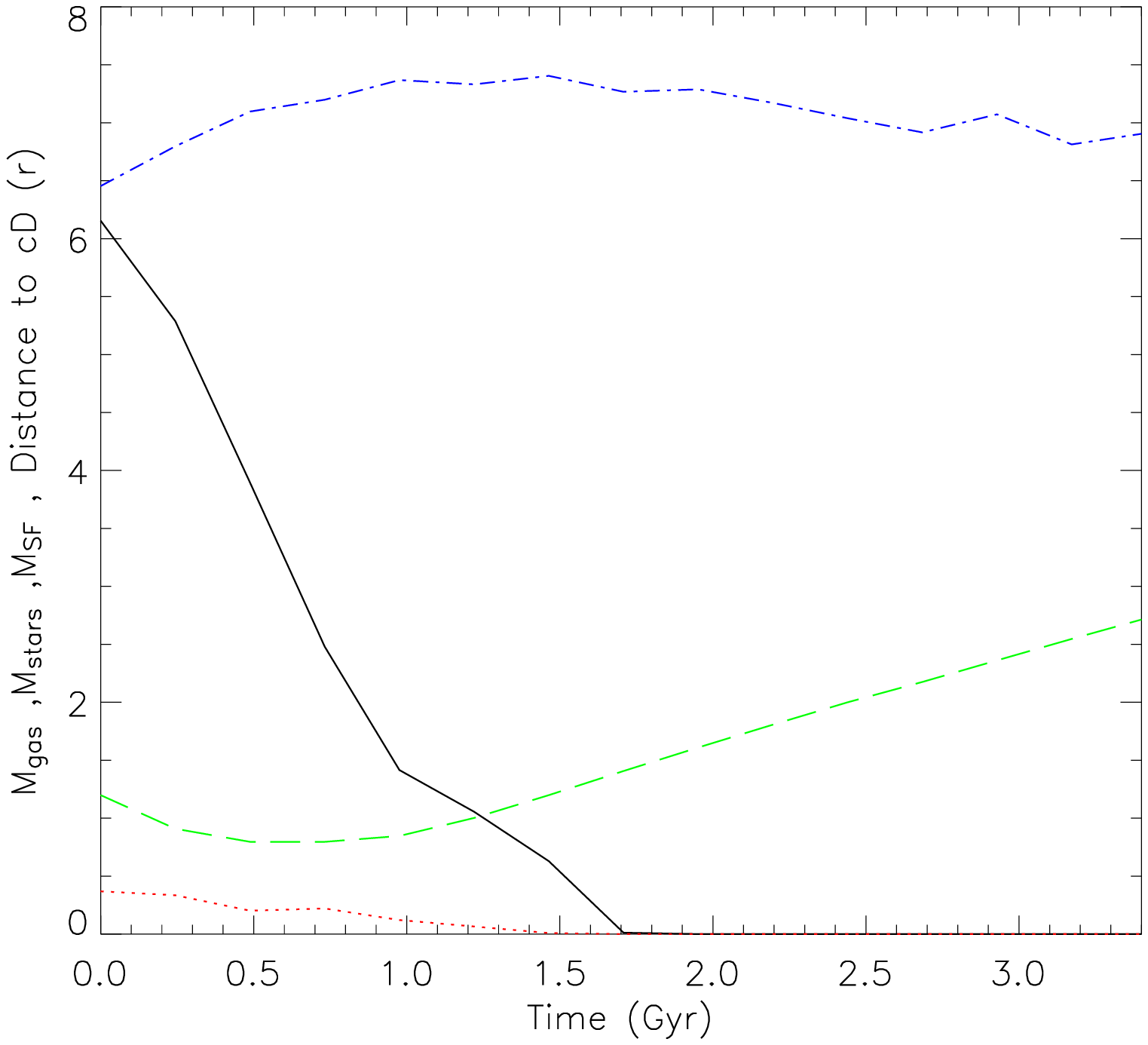}}
\centerline{\plottwo{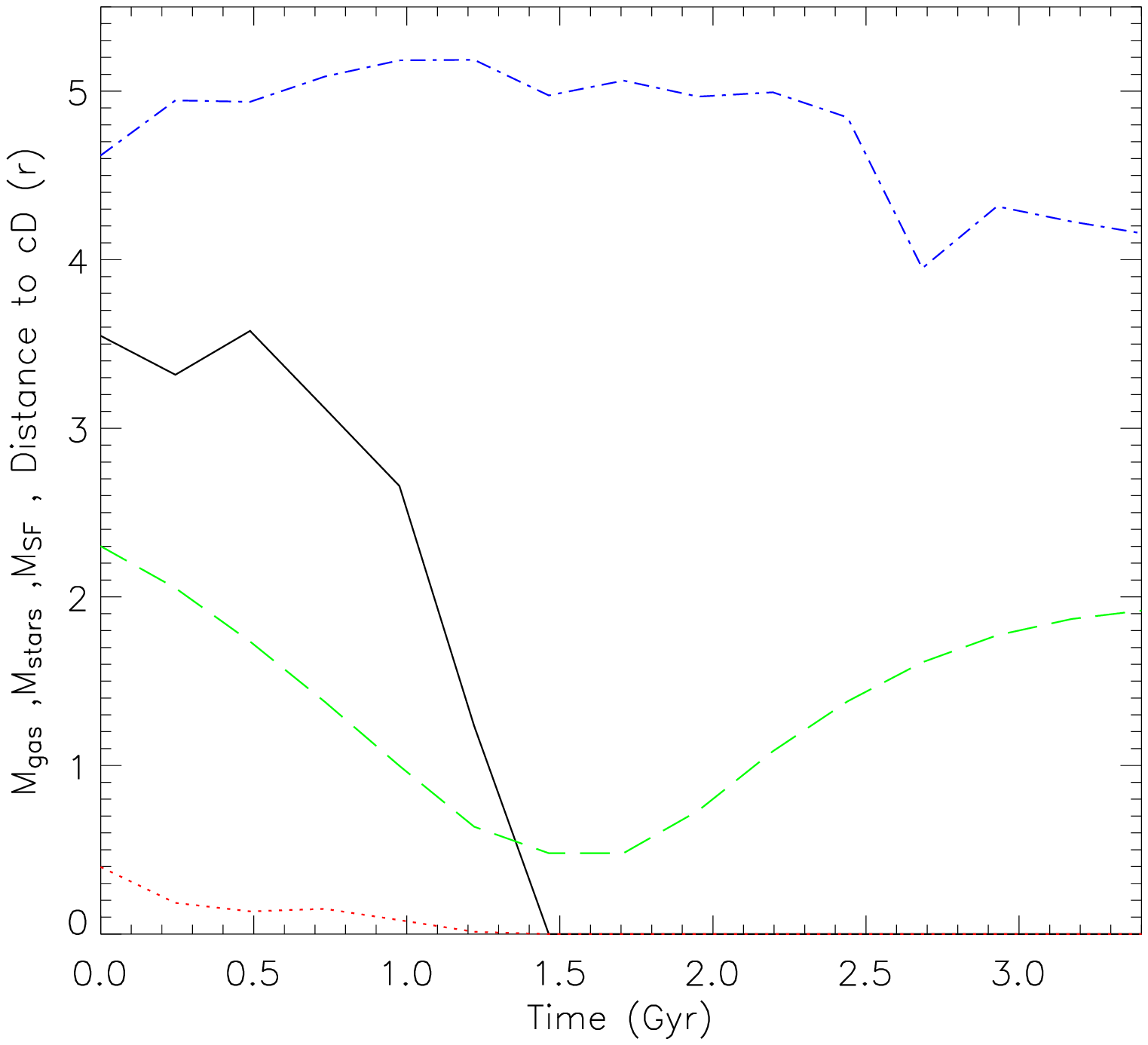}{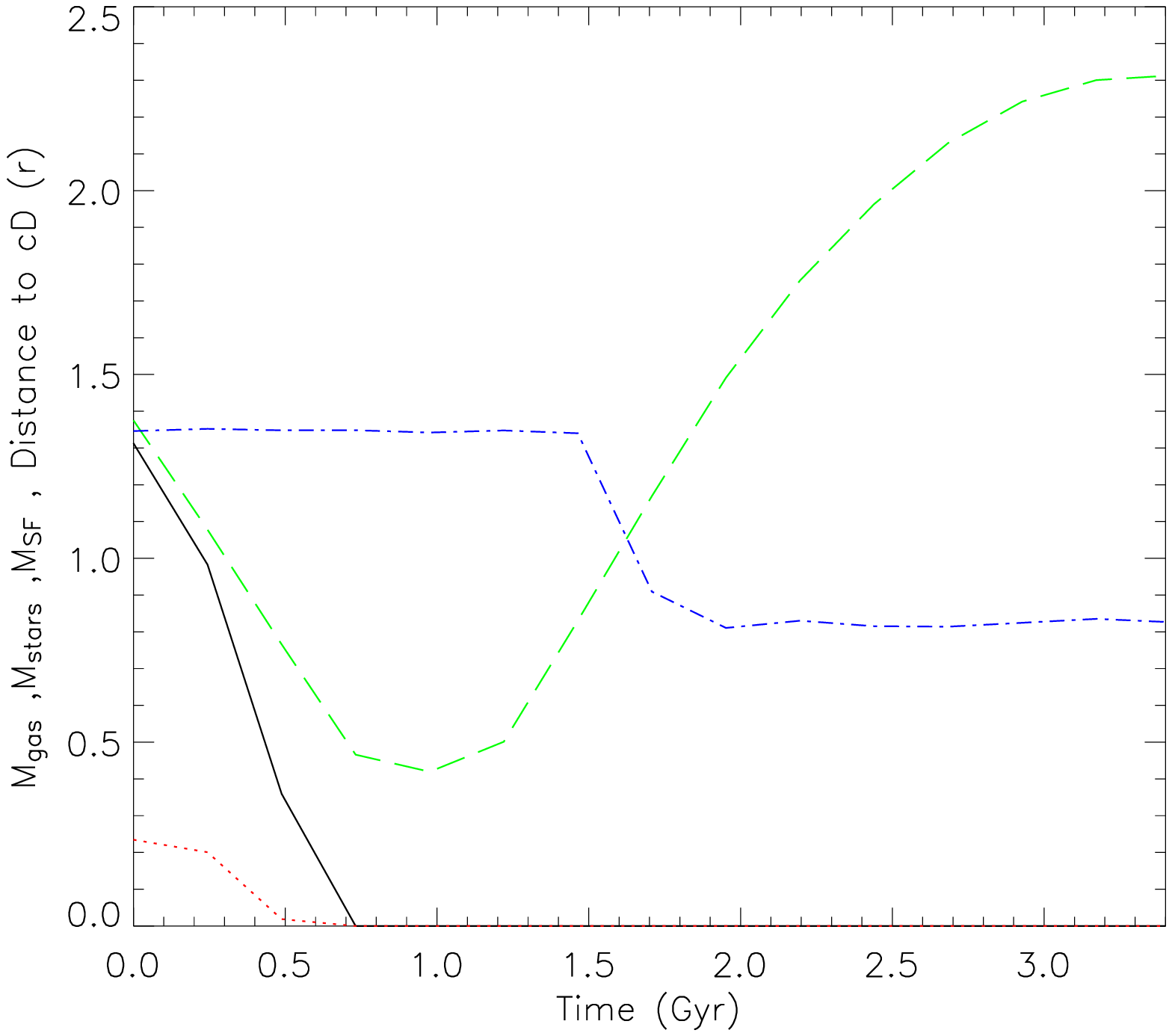}}
\caption{Each graph plots four items against time:  1)  in blue, the total stellar mass of the galaxy (for Fig 7(a) in units of $10^{11} M_{\odot}$, for Fig 7(b-c) in units of $10^{10} M_{\odot}$, and for Fig 7(d) in units of $10^{12} M_{\odot}$, 2) in black, the total gas mass of the galaxy (in units of $10^{10} M_{\odot}$), 3) in red, the amount of mass that will form stars in the next 0.244 Gyr (in units of $10^{10} M_{\odot}$), and 4) in green, the distance from the cD (in Mpc).  The linestyles associated with the four items are shown in the legend.  In this figure, because we also plot the amount of star formation, we do not attempt to make any corrections to the gas or stellar mass to account for it.  See \S 3.4 for discussion. }
\label{fig:evolution}
\end{figure*}

In order to illustrate some of the possible evolutionary paths of the galaxies that lose all their gas, we choose four galaxies to discuss in detail.  In Figure~\ref{fig:evolution}, for each of the galaxies, we plot four quantities as a function of time:  (i) the total stellar mass of the galaxy, (ii) the total gas mass of the galaxy, (iii) the amount of mass that will form stars in the next 0.244 Gyr, and (iv) the distance from the cD.  In this figure, because we also plot the amount of star formation, we do not attempt to make any corrections to the gas or stellar mass to account for it. 

The galaxy in Figure \ref{fig:evolution}a, galaxy A, is most representative of the 16 galaxies we examine in detail, both in terms of mass and evolution.  When we begin to track this galaxy, it has more stellar mass than gas mass, although only by about a factor of three.  Galaxy A's orbit is the most circular of the four chosen galaxies, and we may see most of a circuit of galaxy A around the cD.  This galaxy gains gas in the first 0.244 Gyr of our observations.  Early gas accretion is common in our sample of stripped galaxies, as 56\% of the galaxies that eventually lose all their gas first gain gas for at least one timestep.  It is also not uncommon for a galaxy to gain cool gas mass within the transition region, as this galaxy does.  However, immediately after galaxy A accretes cool gas, its gas is stripped for 0.732 Gyr.  This is one of the faster stripping events we observe.  All the stripping occurs within the central region of the cluster, as predicted by T2003.  Once the galaxy is stripped of its gas (after about 1 Gyr of observations), there are no more significant changes to its mass.

In Figure~\ref{fig:evolution}b, our observations of galaxy B begin when this galaxy has almost the same amount of gas and stellar mass.  This indicates that we may be observing the first time this galaxy has entered the central region of the cluster.  Most of the increase in stellar mass is due to star formation, and we see that it ends when there is no more gas in the galaxy.  All of the small ripples in the stellar mass of the galaxy are too insignificant to be outside of our zero range of ${\pm} 3.16 \times 10^9 M{\odot}$.  For six of the eight timesteps that this galaxy has gas, we categorize it as undergoing ram pressure stripping.  During the other two timesteps the gas mass loss is small and within our zero range.  This is an interesting galaxy because it still has gas when it leaves the cluster center, perhaps because the closest approach is barely within 1 Mpc.  If this galaxy is not anomalous, there should be some galaxies in clusters with displaced gas pointing towards the cD in addition to ram pressure stripped galaxies with tails pointing away from the cD.  One galaxy with a tail pointing towards M86 (the central galaxy in a merging group) in the Virgo cluster has recently been observed by Oosterloo \& van Gorkom (2005).

Galaxy C, in Figure~\ref{fig:evolution}c, also begins with a similar amount of gas and stellar mass. 
The orbital path is consistent with a first entry into the cluster environment, and like galaxy B, the stellar mass increases slightly while the galaxy has gas because of star formation.  As with galaxy A, this galaxy accretes gas before it starts to be stripped.   This galaxy, like half of the galaxies that undergo ram pressure stripping, begins being stripped of gas in the transition region, before it enters the central region of the cluster.  Galaxy C is stripped as far from the cD as 1.7 Mpc.  After 1.5 Gyr this galaxy has no more gas.  Nearly 2.5 Gyr after we begin observing this galaxy it is stripped of a small amount of stars.  At this point the galaxy has passed its closest approach to the cD by almost 1 Gyr and 1 Mpc, so it seems unlikely that material is being stripped by the cD.  Late stellar mass loss is not uncommon:  56\% of the galaxies that become gasless go on to lose stars for at least one timestep.  We speculate that this is due to galaxy harassment.
 
The galaxy in Figure~\ref{fig:evolution}d, galaxy D, is one of the five most massive galaxies we observe in our simulation.  We begin following this galaxy as it falls from the transition region into the central region, however, with our limited amount of orbital information, we cannot tell whether this galaxy is falling towards the central region for the first time.  In this galaxy the amount of stellar mass is two orders of magnitude larger than the amount of gas mass.  The extremely small amount of gas mass leads us to believe that this galaxy has been influenced by the cluster environment for some time.  The gas mass lost in the first 0.244 Gyr is almost entirely due to star formation.  As galaxy D enters the central region of the cluster, a small amount of both gas and stars is lost.  There is another galaxy that we follow that is less than 200 kpc from galaxy D during the second 0.244 Gyr period, and so this may be an example of a galaxy-galaxy tidal stripping event (i.e. harassment).  In the third timestep, there is no change in the stellar mass of galaxy A, but the rest of the gas mass is lost.  Although we measure this as ram pressure stripping, we hesitate to make a definitive categorization because it follows a timestep in which both gas and stars are lost.  Once the gas is stripped from galaxy A, there is no significant change in stellar mass until $\sim1.5$ Gyr into our tracking.  At this point, $43.1 \times 10^{10} M_{\odot}$ is lost in one timestep and $9.86 \times 10^{10} M_{\odot}$ in the next.  Again, the galaxy has passed its closest approach to the cD (by almost 0.5 Gyr and 0.5 Mpc).  After this $\sim0.5$ Gyr stellar stripping event, the galaxy undergoes no more significant mass changes.

There are a few important points that this subsample of galaxies highlights.  First, approximately 75\% of the galaxies that lose all their gas are affected by ram pressure stripping, often losing most of their gas by this mechanism.  In half of the ram pressure stripped galaxies, the stripping begins in the transition region, further than assumed by T2003, although most of the gas is lost in the central region.  As discussed in the introduction, there have been observations of ram pressure stripping far from the cluster center.  Gas stripping tends to be a long process, generally taking at least 1 Gyr.  Also, once these galaxies lose their gas, over half of them undergo a stellar stripping event that is not clearly due to the cD.

\section{Conclusions}
\label{sec:conclusions}

In this paper we have presented a first examination of how the gas and stellar content of galaxies evolve within a cosmological simulation of a cluster of galaxies.  We use a high resolution simulation that includes the required gas, dark matter and stellar physics in order to find out how and when galaxies lose their mass.  Our main results are:

\begin{enumerate}

\item  We have tracked 132 galaxies through time in a detailed simulated cluster environment.  We make comparisons with recent observations, specifically those of Treu et al (2003).  Like T2003 we split our cluster into three regions:  central region ($r < 1$ Mpc), transitional region ($1 < r < 2.4$ Mpc), and the periphery ($r > 2.4$ Mpc).  We find a relation between the cluster-radius and galaxy gas content that is qualitatively similar to that found by T2003 (see Table~\ref{tbl-1}), although we note that a detailed comparison is difficult by our inability to assign a reliable morphological class to our simulated galaxies.  

\item Most of the gas lost from galaxies in our simulations is lost in a gas-only event (i.e. the stellar mass in unchanged).  These events are preferentially found in the central region, but can occur as far out as 2 Mpc from the cluster center.  We find that the amount of gas loss correlates with the ram-pressure experienced by the galaxy, indicative of a ram-pressure origin to the gas loss.  At fixed radius from the cluster, there is a wide variation in the ram-pressure strength experienced by a given galaxy.

\item  We observe mergers both in the central region of our cluster as well as in the periphery, consistent with T2003.  We do not observe any dry mergers (although they might occur in the galaxies we do not follow), and none of the mergers we follow exhaust the gas supply of the participating galaxies.  Further, we observe disruptions of both the gas and stellar mass in galaxies in all three regions that could be attributed to galaxy harassment or other galaxy-galaxy interactions (Figure~\ref{fig:mass_change} \& Table~\ref{table:changes}).

\item Galaxies in the periphery and field ($r > 2.4$ Mpc) are observed to accrete cold gas; however, this accretion is largely suppressed for galaxies in the transition and central regions ($r < 2.4$ Mpc).  We interpret this as starvation caused by the ICM.

\item  By examining in detail the galaxies that lose all their gas, focusing on four different cases in particular, we were able to draw more detailed conclusions about this small subset of 16 galaxies.  First, ram pressure stripping, which affected at least 12 of these galaxies, is the dominant mechanism causing galaxies to lose their gas.  Ram pressure stripping began in the transition region for half of the stripped galaxies, and in one case in the periphery.  In agreement with T2003, we find that gas stripping tends to occur on timescales $\ge$1 Gyr.  Although these total gas stripping events may begin as starvation, only effecting the outer halo gas associated with these galaxies, they clearly end by removing any gas that would have been in the galactic disk.  As addressed above, we cannot make any claims about the fate of dense molecular gas.  We also found that many galaxies, once they lost their gas, also lost a significant amount of stellar mass.  It was not clearly correlated with a galaxy's closest approach to the cD, nor was it followed by a merger.  This finding may lend tentative support to galaxy harassment as an important mass stripping mechanism.

\end{enumerate}

We interpret these results to mean that the decrease in gasless galaxy fraction with increasing cluster radius can be explained by environmental mechanisms out to almost 2 Mpc.  This result parallels observational findings by, eg, Solanes et al (2001).  ISM-ICM interactions are important out to this large radius, and ram pressure stripping may have a large role in transforming spirals into S0s out to this distance. The ICM in our simulation has significant substructure, which can been seen in the spread of ram pressure values at any cluster radius in Figure~\ref{fig:ram_distance}.  A similar range of ICM density at different clustercentric radii is seen in our simulations, and density variations have also been observed (e.g. Bohringer et al. 1994).  The ICM's structure could explain why it is more important than in the simple assumptions used by T2003.  However, the stripping process can be very slow ($\ge$ 1 Gyr), and therefore conforms to the broad definition of starvation used by T2003.

We make clear predictions about RPS, and can compare the mass evolution caused by galaxy-galaxy and galaxy-cluster interactions with that caused by ISM-ICM interactions.  However, we do not compare galaxy-galaxy and galaxy-cluster interactions.  This is because these interactions can have the same signature effects on the gas and stellar mass of a particular galaxy.  In order to make any comparisons we will have to make detailed calculations about the force over time of nearby galaxies and the cD.  Also, although we see definite trends with radius, we have not begun to look at whether there is a relation between the local density of galaxies and evolutionary mechanism.  These will wait for a future examination. 

\acknowledgments

We gratefully acknowledge the National Center for Supercomputing Applications, which provided computational resources for the simulation described in this paper.  Greg Bryan acknowledges support from NSF grants AST-05-07161, AST-05-47823, and AST-06-06959. JvG acknowledges support from NSF grant AST-06-07643.

\appendix

\section{Resolution Study}

We have performed a set of more detailed single-galaxy simulation runs to verify that our results are resolution independent.  We use the galaxy model of Roediger {\&} Br{\"u}ggen (2006), with the addition of radiative cooling (but no star formation).  The ICM temperature and density are $4.385 \times 10^7$ K and $10^{-28}$ g cm$^{-3}$, respectively.  The two runs we present here have resolutions of 304 pc and 2.43 kpc.  As necessitated by the different resolutions, the gas disk scale height in the z direction increases from 0.4 kpc to 4.0 kpc.  We perform runs with two velocities, a subsonic and supersonic case:  $8.0 \times 10^7 cm s^{-1}$ and $2.53 \times 10^8 cm s^{-1}$.  Unlike Roediger {\&} Br{\"u}ggen (2006), all of the ICM in our simulation instantaneously begins to move at the wind speed.  As in our paper, we follow the cool (T $\le 15,000$ K) gas mass within a sphere with a radius of 26.7 kpc.  Although we start with no gas cooler than 15,000 K, most of the gas within the galaxy quickly cools to below our upper limit.  

\begin{figure}
\centerline{\includegraphics[scale=0.9]{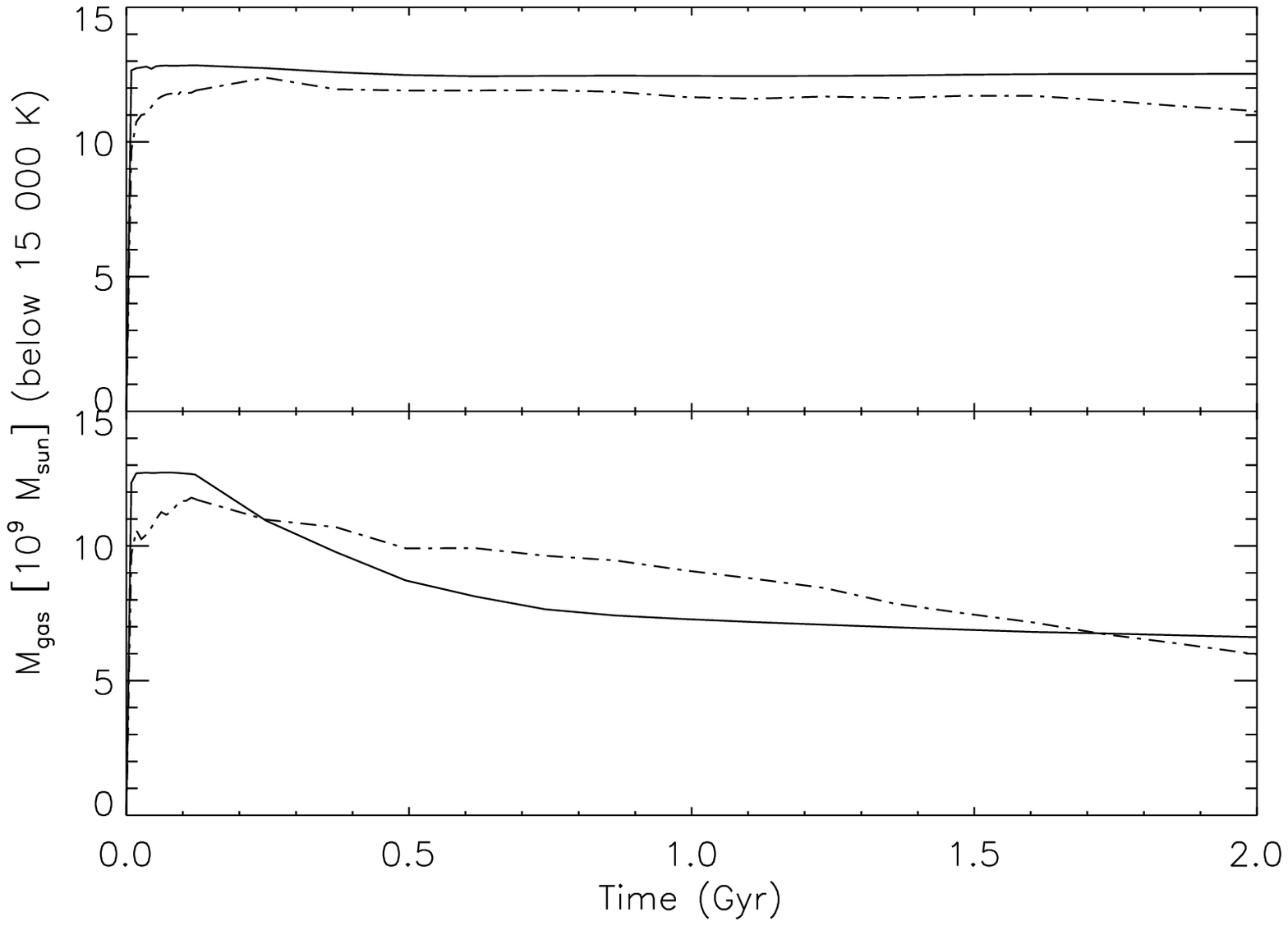}}
\caption{The gas mass evolution of galaxies with different resolutions in a subsonic (upper panel) and supersonic (lower panel) wind.  The solid line is a galaxy with 304 pc resolution based on the model Roediger \& Br{\"u}ggen (2006), while the dash-dot line is a galaxy with 2.43 kpc resolution based on the same model but with a z scale height ten times larger.  The difference in gas loss is small enough that we are confident that our resolution is sufficient for following gas loss.}
\label{fig:a1}
\end{figure}

We show our results in Figure~\ref{fig:a1}.  As seen in the upper panel, very little gas is lost in either of the galaxy models in the subsonic run.  In the supersonic run shown in the lower panel, the galaxy with the smaller scale height and higher resolution initially loses cool gas more quickly.  However, the disk with higher resolution keeps a smaller disk of cool gas, while the lower resolution disk continues to slowly be stripped of it's gas with time.  Because we do not include star formation, we are missing the energy that would be input by the resulting supernovae and increase the height of the disk.  We perform a simulation without radiative cooling, and therefore with a thicker disk, and find that the gas loss between galaxies of different resolutions is more similar.  We expect that including star formation results in disks with larger z scale heights, and therefore that the gas loss measured in the galaxies in the cosmological simulation is less effected by resolution differences than in the galaxies we show here (with only radiative cooling).  Similar results are found when the galaxy is edge-on to the wind.  As shown in Figure~\ref{fig:a1}, although the gas loss history differs in the two models, the difference is never large.  Based on these results we are confident that our resolution is high enough to measure the amount of cool gas a galaxy may lose outside of a small dense disk.

\end{document}